\documentclass[onecolumn,draftcls,journal]{IEEEtran}

\usepackage{graphicx}
\usepackage{amsfonts}
\usepackage{amssymb}
\usepackage{psfrag}
\usepackage{pstricks}
\usepackage[intlimits,sumlimits]{amsmath}
\usepackage{ifthen}
\usepackage{times}
\usepackage{latexsym}
\usepackage{setspace}

\newtheorem{theorem}{Theorem}
\newtheorem{corollary}{Corollary}

\newcommand {\exe} {\stackrel{\cdot} {=}}
\newcommand {\reals} {{\rm I\!R}}
\newcommand{\bm}[1]{\mbox{$\bf #1$}}

\newcommand{\calV}{{\cal V}}

\newcommand{\calX}{{\cal X}}
\newcommand{\calY}{{\cal Y}}

\begin{document}

\title{Asymptotically Optimum Universal One--Bit
Watermarking for Gaussian Covertexts and Gaussian Attacks}

\author{Pedro Comesa{\~n}a, Neri Merhav,
and Mauro Barni
\thanks{P. Comesa{\~n}a is with the Signal Theory and Communications
Department, University of Vigo, Campus
Lagoas-Marcosende, Vigo 36310, Spain, (phone: +34 986 812683,
fax: +34 986 812116, e-mail:
pcomesan@gts.tsc.uvigo.es),
N. Merhav is with the Department of Electrical Engineering,
Technion -- I.I.T., Haifa 32000, Israel, (phone/fax: +972-4-8294737, e-mail:
merhav@ee.technion.ac.il),
M. Barni is with the Department of Information Engineering,
University of Siena, Via Roma 56, Siena 53100, Italy, (phone: +39
0577 234624 / +39 0577 234621, fax: +39 0577 233630, e-mail:
barni@dii.unisi.it).}
\thanks{This work
was partially supported by the Italian Ministry of Research and
Education under FIRB project no. RBIN04AC9W.}}

\maketitle

\begin{abstract}
The problem of optimum watermark embedding and detection was
addressed in a recent paper by Merhav and Sabbag, where the
optimality criterion was the maximum false--negative error exponent
subject to a guaranteed false--positive error exponent. In
particular, Merhav and Sabbag derived universal asymptotically
optimum embedding and detection rules under the assumption that the
detector relies solely on second order joint empirical statistics of
the received signal and the watermark. In the case of a Gaussian
host signal and a Gaussian attack, however, closed--form expressions
for the optimum embedding strategy and the false--negative error
exponent were not obtained in that work. In this paper, we derive
such expressions, again, under the universality assumption that
neither the host variance nor the attack power are known to either
the embedder or the detector. The optimum embedding rule turns out
to be very simple and with an intuitively--appealing geometrical
interpretation. The improvement with respect to existing
sub--optimum schemes is demonstrated by displaying the optimum
false--negative error exponent as a function of the guaranteed
false--positive error exponent.
\end{abstract}

\begin{IEEEkeywords}
Watermarking, watermark embedding, watermark detection, hypothesis testing, Neyman--Pearson.
\end{IEEEkeywords}

\clearpage

\section{Introduction}
\label{sec:Introduction}

About a decade ago, the community of researchers in the field of
watermarking and data hiding
has learned about the importance and relevance of
the problem of channel coding with non--causal side information at the transmitter
\cite{GP80}, and in particular, its Gaussian version -- {\it writing on dirty paper}, due to
Costa \cite{Costa83}, along with its direct applicability to watermarking, cf.\
\cite{Cox99,Chen01}. Costa's main result is that the capacity of the
additive white Gaussian noise (AWGN) channel with an additional
independent interfering signal, known non--causally to the
transmitter only, is the same as if this interference was
available at the decoder as well (or altogether non--existent). When
applied in the realm of watermarking and data hiding, this means
that the host signal (playing the role of the interfering signal),
should not be actually considered as additional noise, since the
embedder (the transmitter) can incorporate its knowledge upon
generating the watermarked signal (the codeword). The methods based
on this paradigm, usually known as {\em side-informed} methods, can
even asymptotically eliminate (under some particular conditions) the
interference of the host signal, that was previously believed to be
inherent to any watermarking system.

Ever since the relevance of Costa's result to watermarking has been
observed, numerous works have been published about the practical
implementation of the side--informed paradigm for the so-called {\it
multi--bit watermarking}
\cite{Chen01,Ramkumar04,Abrardo05,Fernando05} case, where the
decoder estimates the transmitted message among many possible
messages. Far less attention has been devoted, however, to the
problem of deciding on the presence or absence of a given watermark
in the observed signal. In fact, in most of the works that deal with
this binary hypothesis testing problem, usually known as one--bit
(a.k.a.\ zero--bit) watermarking, the watermarking displacement
signal does not depend on the the host
\cite{Her00,derosa,Huang07,Noorkami07,Dong07} that then interferes
with the watermark, thus contributing to augment the error
probability. To the best of our knowledge, exceptions to this
statement are the works by Cox {\it et al.} \cite{Cox99,Miller00},
Liu and Moulin \cite{Liu03}, Merhav and Sabbag \cite{Merhav08} and
Furon \cite{Furon07}. In the next few paragraphs, we briefly
describe the main results contained in these works.

\vspace{0.25cm}

\noindent
{\em Cox et. al. \cite{Cox99,Miller00}}:
In \cite{Cox99}, Cox {\it et. al.} introduce the
paradigm of watermarking as a coded communication system with side
information at the embedder. Based on this paradigm, and by
considering a statistical model for attacks, the
authors propose a detection rule based on the Neyman--Pearson criterion.
The resulting detection region is replaced by the union of
two hypercones; mathematically, this detection rule is given by
$\frac{ |\bm s^t \cdot \bm u|}{\|\bm
s\|\cdot\| \bm u\|} \geq \tau(\alpha)$, where $\bm s$ is the received
signal, $\bm u$ is the watermark, $\bm s^t$ is the transpose of $\bm s$,
$\bm s^t \cdot \bm u$ is the inner product of $\bm s$ and $\bm u$,
$\alpha$ is the maximum allowed false--positive probability, and
$\tau(\alpha)$ is the decision threshold, which is a function of
$\alpha$. In a successive paper \cite{Miller00}, Miller {\it et al.}
also compare the performance of the strategy of \cite{Cox99}
to other typical embedding strategies. No attempt is made to jointly
design the optimum embedding and detection rules.

\vspace{0.25cm}

\noindent
{\em Liu and Moulin \cite{Liu03}}:
In \cite{Liu03}, both
false--positive and false--negative error exponents are studied for the
one--bit watermarking problem, both for additive spread spectrum
(Add-SS) and a quantization index modulation (QIM) technique
\cite{Chen01}. The constraint on the embedding
distortion is expressed in terms of the mean Euclidean norm of the
watermarking displacement signal, and the non--watermarked signal is also
assumed to be attacked (with attacks that impact the
false--positive error probability). For Add-SS, exact expressions
of the error exponents of both false--positive and false--negative
probabilities are derived. For QIM, the authors provide
bounds only. These results show
that although the error exponents of QIM are indeed larger than
those obtained by public Add-SS (where the host signal is not
available at the detector), they are still smaller than those
computed for private Add-SS (where the host signal is also available
at the detector). This seems to indicate that the interference
due to the host is not completely removed.

\vspace{0.25cm}

\noindent {\em Merhav and Sabbag \cite{Merhav08}}: In
\cite{Merhav08}, the problem of one--bit watermarking is approached
from an information--theoretic point of view. Optimum embedders and
detectors are sought, in the sense of minimum false--negative
probability subject to the constraint that the false--positive
exponent is guaranteed to be at least as large as a given prescribed
constant $\lambda> 0$, under a certain limitation on the kind of
empirical statistics gathered by the detector. Another feature of
the analysis in \cite{Merhav08} is that the statistics of the host
signal are assumed unknown. The proposed asymptotically optimum
detection rule compares the empirical mutual information between the
watermark $\bm u$ and the received signal $\bm y$ to a threshold
depending on $\lambda$. In the Gaussian case, this boils down to
thresholding the absolute value of the empirical correlation
coefficient between these two signals. Merhav and Sabbag also derive
the optimal embedding strategy for the attack--free case and derive
a lower bound on the false--negative error exponent. Furthermore,
the optimization problem associated with uptimum embedding is
reduced to an easily implementable 2D problem yielding a very simple
embedding rule. In that paper, Merhav and Sabbag study also the
scenario where the watermarked signal is attacked. In this case,
however, closed--form expressions for the error exponents and the
optimum embedding rule are not available due to the complexity of
the involved optimizations.

\vspace{0.25cm}

\noindent
{\em Furon \cite{Furon07}}:
In \cite{Furon07}, Furon uses the Pitman--Noether theorem \cite{Poor}
to derive the form of the best detector for a given embedding
function, and the best embedding function for a given detection
function. By combining these results, a differential equation is
obtained, that the author refers to as the {\em fundamental equation
of zero-bit watermarking}. Furon shows that many of the most
popular watermarking methods in the literature can be seen as
special cases of the fundamental equation, ranging from
Add-SS, multiplicative spread spectrum, or JANIS
\cite{Furon02JANIS}, to a two-sheet hyperboloid, or even
combinations of the previous techniques with watermarking on a
projected domain \cite{Bala03}, or watermarking based on lattice
quantization. Compared with the framework introduced in \cite{Merhav08},
two important differences must be highlighted:
\begin{itemize}
\item In \cite{Furon07}, the watermarking displacement signal is constrained to be
a function of the host signal which is scaled to yield a given
embedding distortion. This means that in this set--up the direction of the
watermarking displacement signal
can not be changed as a function of the allowed embedding distortion.
\item One of the conditions that must be verified in order to apply
the Pitman--Noether theorem is that the power of the
watermarking displacement signal goes to zero when the dimensionality increases without bound.
In fact, Furon hypothesizes that this is the reason why neither
the absolute normalized correlation nor the normalized correlation
are solutions of the fundamental equation.
\end{itemize}

In this paper, we extend the results of \cite{Merhav08} and derive a
closed--form expression for the optimum embedding and detection strategies in the
Gaussian set--up, that is, for a Gaussian host signal and a Gaussian
attack channel. As in \cite{Merhav08}, we assume that the embedder
and the detector do not know the variance of the host signal and
that of the noise added by the attacker. We also share with
\cite{Merhav08} the assumption that the detector is
of limited resources, specifically, that it relies only on the Euclidean
norm of the received signal and the empirical correlation between the received
signal and the watermark. We
derive explicit embedding and detection rules and establish their
asymptotic optimality
in the Neyman--Pearson sense of maximizing the
false--negative error exponent
for a given guaranteed false--positive error exponent.
We also derive a closed--form expression for the false--negative error exponent.
The optimum embedding strategy
turns out to be very simple, and this opens the door to the
development of new practical watermarking schemes for real--life signals
like images, video or audio signals. The improved performance of the
new scheme is demonstrated both theoretically, by comparing the
achieved error exponents and those achieved by previous methods, and
numerically, by displaying graphs of the error exponent functions.

The remaining part of the paper is organized as follows: In
Section~\ref{sec:notation}, we introduce notation conventions and formalize the problem.
In Section~\ref{sec:opt_region}, an asymptotically optimum detection region is derived.
In Section~\ref{sec:false_negative}, we use it to compute the false--negative error
exponent, whose optimization is considered in
Section~\ref{sec:opt_emb} to derive a corresponding optimum embedder.
In Section~\ref{sec:noiseless}, the optimum embedder and
the exact false--negative error exponent for the noiseless case are
introduced as a by--product of this analysis and compared to
previous results in the literature. Finally, the main
results of this work are summarized in Section~\ref{sec:conclusions}
where some suggestions for future research are also outlined.

\section{Notation and Problem Formulation}
\label{sec:notation}

Throughout the sequel, we denote scalar random variables by capital letters
(e.g., $V$), their realizations with corresponding lower case letters (e.g., $v$), and
their alphabets, with the respective script font (e.g., $\calV$).
The same convention applies to $n$--dimensional random vectors and
their realizations, using bold face fonts (e.g., $\bm V$,
$\bm v$). The alphabet of each corresponding $n$--vector will be taken to be
the $n$--th Cartesian power of the alphabet of a single component, which
will be denoted by the alphabet of a single component with a superscript $n$ (e.g.,$\calV^n$).
The $i$-th component of a vector $\bm V$ is denoted
$V_i$. The probability law of a random vector $\bm V$ is
described by its probability density function (pdf) $f_{\bm V}(\bm
v)$, or its probability mass function (pmf) $P_{\bm V}(\bm V = \bm
v)$, depending on whether it is continuous or discrete, respectively.

Let $\bm u$ and $\bm x$, both $n-$dimensional vectors,
be the {\it watermark sequence} and the {\it host sequence},
respectively. While $u_i$, $i=1,\ldots,n$,
the components of $\bm u$, take on binary values in
${\cal U}=\{-1,+1\}$,\footnote{The basic derivations of this work
will remain valid for different
choices of ${\cal U}$.}
the components of $\bm x$, namely, $x_i$, $i=1,\ldots,n$, take
values in $\calX=\reals$.
The embedder receives $\bm x$ and $\bm u$, and
produces the {\it watermarked sequence} $\bm y$, yet another
$n$--dimensional vector with components in $\calY=\reals$. We refer to
the difference signal $\bm w=\bm y-\bm x$ as the {\it watermarking displacement signal}.
The embedder must keep the
embedding distortion $d(\bm x, \bm y) =\|\bm y-\bm x\|^2$
within a prescribed limit, i.e., $d(\bm x, \bm y) \leq n D$,
where $D>0$ is the maximum allowed distortion per dimension, uniformly
for every $\bm x$ and $\bm u$.

The output signal of the transmitter
may either be the unaltered original host $\bm x$, in the
non--watermarked case, or the vector $\bm y$, in the watermarked case.
In both cases, this output signal is subjected to an attack, which yields
a {\it forgery} signal, denoted by $\bm s$.
The action of the attacker is modeled by a channel, which is given in terms of a
conditional probability density
of the forgery given the input it receives,
$W(\bm s|\bm x)$ -- in the non--watermarked case,
or $W(\bm s|\bm y)$ -- in the watermarked case.
For the sake of convenience, we
define $\bm z$ as the noise vector added by the attacker, i.e., the difference between
the forgery signal $\bm s$ and the channel input signal, which is the transmitter output
($\bm x$ or $\bm y$, depending on whether the signal is watermarked or not).
We assume that $\bm z$ is a Gaussian vector with zero--mean, i.i.d.\ components, all having
variance~$\sigma_Z^2$.

The detector partitions $\reals^n$ into two complementary regions,
$\Lambda$ (a.k.a.\ the detection region) and $\Lambda^c$. If $\bm
s\in\Lambda$, the detector decides that the watermark is present,
otherwise it decides that the watermark is absent. We assume that
the detector knows the watermark $\bm u$, but does not know the host
signal $\bm x$ (blind or public watermarking). The design of the
optimum detection region for the attack--free case was studied in
\cite{Merhav08}, and it is generalized to the case of Gaussian
attacks in Section~\ref{sec:opt_region}.

The performance of a one--bit watermarking system is usually
measured in terms of the tradeoff between the {\em false positive}
probability of deciding that the watermark is present when it is
actually absent, i.e.,
\begin{equation}
\label{pfp}
P_{fp}=\int_{\Lambda}\mbox{d}\bm s \cdot[2\pi(\sigma_X^2+\sigma_Z^2)]^{-n/2}\cdot
\exp\left\{-\frac{\|\bm s\|^2}{2(\sigma_X^2+\sigma_Z^2)}\right\}
\end{equation}
and the {\em false negative}
probability, of deciding that the watermark is absent when it is actually present, i.e.,
\begin{equation}
\label{pfn}
P_{fn}=\int_{\Lambda^c}\mbox{d}\bm s \int_{\reals^n}\mbox{d}\bm x\cdot
(2\pi\sigma_X^2)^{-n/2}\cdot\exp\left\{-\frac{\|\bm x\|^2}{2\sigma_X^2}\right\}\cdot
(2\pi\sigma_Z^2)^{-n/2}\cdot\exp\left\{-\frac{\|\bm s-f(\bm x,\bm u)\|^2}{2\sigma_Z^2}\right\},
\end{equation}
where $f$ is the embedding function, that is, $\bm y=f(\bm x,\bm u)$. As $n$ grows
without bound, these probabilities normally decay exponentially. The
corresponding exponential decay rates, i.e., the {\em error
exponents}, are defined as
\begin{equation}
\label{efp}
  E_{fp} \triangleq \lim_{n\rightarrow \infty} - \frac{1}{n}\ln P_{fp},
\end{equation}
\begin{equation}
\label{efn}
  E_{fn} \triangleq \lim_{n\rightarrow \infty} - \frac{1}{n}\ln P_{fn}.
\end{equation}

The aim of this paper is to devise a detector as well as an
embedding rule for a zero--mean, i.i.d.\ Gaussian host with variance
$\sigma_X^2$ and a zero--mean memoryless Gaussian attack channel
with noise power $\sigma_Z^2$, where the detector is limited to base
its decision on the empirical energy of the received signal and its
empirical correlation with $\bm u$. Both $\sigma_X^2$ and
$\sigma_Z^2$ are assumed unknown to the embedder and the detector.
We seek optimum embedding and detection rules in the sense of
uniformly maximizing the false--negative error exponent, $E_{fn}$,
(across all possible values of $\sigma_X^2$ and $\sigma_Z^2$)
subject to the constraint that $E_{fp}\ge \lambda$, where $\lambda$
is a prescribed positive real.

\section{Optimum Detection and Embedding}\label{sec:opt_region}

In \cite{Merhav08}, an asymptotically optimum detector is derived
for the discrete case and for the continuous Gaussian case. In the
latter case, it is shown that if the detector is limited to base its
decision on the empirical energy of the received signal,
$\frac{1}{n}\sum_{i=1}^ns_i^2$, and its empirical correlation with
the watermark, $\frac{1}{n}\sum_{i=1}^nu_is_i$, then an
asymptotically optimum decision strategy, in the above defined
sense, is to compare the (Gaussian) empirical mutual information,
given by:
\begin{eqnarray}
  \hat{I}_{\bm u\bm s}(U;S) = - \frac{1}{2} \ln \left[1 - \frac{\left(\frac{1}{n}
        \sum_{i=1}^n u_i s_i \right )^2}
   { \left ( \frac{1}{n}\sum_{i=1}^n u_i^2 \right )\left(\frac{1}{n}\sum_{i=1}^n s_i^2 \right )} \right]
   = - \frac{1}{2} \ln \left[1 - \frac{\left(\frac{1}{n}
        \sum_{i=1}^n u_i s_i \right )^2}
   { \frac{1}{n}\sum_{i=1}^n s_i^2} \right]
   \label{eq:detector}
\end{eqnarray}
to $\lambda$, or equivalently, to compare
the absolute normalized correlation
\begin{eqnarray}
\label{eq:rho-det}
|\hat{\rho}{\bm u \bm s}| = \frac{\left|\frac{1}{n}
        \sum_{i=1}^n u_i s_i \right|}
   {\sqrt{\frac{1}{n}\sum_{i=1}^n s_i^2}}, \label{eq:detector2}
\end{eqnarray}
to $\sqrt{1-e^{-2\lambda}}$, i.e., the detection region is the union of two hypercones,
around the vectors $\bm u$ and $-\bm u$, with a spread depending on $\lambda$. This decision rule
of thresholding the empirical mutual information, or empirical correlation, is intuitively appealing
since the empirical mutual information is an estimate of the degree of statistical dependence
between two data vectors.\footnote{It is also encountered in the literature of universal decoding
the maximum mutual information (MMI) decoder
for unknown memoryless channels.}

For the present setting, we have to extend the analysis
to incorporate the Gaussian attack channel.
But this turns out to be staightforward, as in the non--watermarked
case (pertaining to the false--positive constraint), $\bm s$ continues to be Gaussian -- the only
effect of the channel is in changing its variance, which is assumed unknown anyhow.
Thus, the same detection rule as above continues to be asymptotically optimum in our setting as well.

Before we proceed to the derivation of the optimum embedder,
it is instructive to look more closely at the dependence
of the detection region on the false--positive exponent $\lambda$. As
mentioned earlier, the choice of $\lambda$ imposes
a threshold that must be compared with (\ref{eq:detector2}) in order
to provide the detector output. This is equivalent to establishing
the limit angle of the detection region, that we will denote by
$\beta=\arccos(\sqrt{1-e^{-2\lambda}})=\arcsin(e^{-\lambda})\in[0,\pi/2]$.
Letting $\theta=\arccos(\hat{\rho}_{\bm u\bm s})$, we then have:
\begin{eqnarray}
P_{fp} &=& \textrm{Pr}\{\hat{\rho}_{\bm u\bm s}^2 > 1-e^{-2\lambda}|H_0\}\nonumber\\
&=& \textrm{Pr}\{0 \leq \theta < \beta |H_0\} +
\textrm{Pr}\{\pi - \beta < \theta \leq \pi |H_0\} \nonumber \\ &=&
2 \textrm{Pr}\{0 \leq \theta < \beta |H_0\} = \frac{2A_n(\beta)}{A_n(\pi)}
  \doteq e^{n \ln(\sin\beta)},\label{eq:detec_fp}
\end{eqnarray}
where the notation $\doteq$ stands for equality
in the exponential scale as a function of $n$,\footnote{
More precisely, if $\{a_n\}$ and $\{b_n\}$ are two positive sequences, $a_n\exe b_n$ means
that $\lim_{n\to\infty}\frac{1}{n}\log\frac{a_n}{b_n}=0$.}
and where $A_n(\theta)$ is the surface area of the $n$--dimensional
spherical cap cut from a unit sphere centered in the origin, by a
right circular cone of half angle $\theta$.
In (\ref{eq:detec_fp}), we used the fact
that in the non--watermarked case, where $\bm s$ is a zero--mean Gaussian vector
with i.i.d.\ components, independent of $\bm u$,
the normalized vector $\bm s/\|\bm s\|$ is
uniformly distributed across the surface of the $n$--dimensional
unit sphere, as there are no preferred directions.
The exact expression of $A_n(\theta)$ is given by:
\begin{eqnarray}
  A_n(\theta) =\frac{(n-1) \pi^{(n-1)/2}}{\Gamma\left (\frac{n+1}{2} \right)}
  \int_{0}^\theta \sin^{(n-2)}(\varphi) d\varphi. \nonumber
\end{eqnarray}

\section{The False--Negative Exponent}\label{sec:false_negative}

In this section, we make the first step towards the derivation of the
optimum embedding strategy. In particular, we compute
the false--negative error exponent as a function of the watermarking displacement
signal $\bm w$, which is represented by a three--dimensional vector
$\bm v=(v_1,v_2,v_3)$. The vector $\bm v$ is
the vector $\bm w$, normalized by $\sqrt{n}$, and transformed to
the coordinate system pertaining to the
linear subspace spanned by $\bm u$, $\bm x$ and $\bm w$.
This result will be used later to derive the optimal
embedding function subject to the
distortion constraint, that limits
the norm of $\bm w$ not to exceed $nD$, which corresponds to the constraint
$v_1^2+v_2^2+v_3^2\le D$.
To this end, we establish the following theorem.

\begin{theorem}\label{th:teorema}
Let $P_{fp}$, $P_{fn}$ and their corresponding error exponents $E_{fp}$ and $E_{fn}$, be defined as in
eqs.\ (\ref{pfp}),(\ref{pfn}),(\ref{efp}) and (\ref{efn}), respectively.
Let $\bm v=(v_1,v_2,v_3)\in\reals^3$ be given,
and let $\Lambda=\{\bm s:~\hat{\rho}_{\bm u\bm s}^2\ge 1-e^{-2\lambda}\}$. Then,
  \begin{eqnarray}
    E_{fn} &=&
\min_{q\in[\max(0, T_1(r,\alpha,\bm v)), \infty)}\min_{r\in[0,\infty)}\min_{\alpha\in[-\pi/2,\pi/2]}
\left\{\frac{1}{2} \left[\frac{q}{\sigma_Z^2} - \ln \left (\frac{q}{\sigma_Z^2}
\right ) - 1 \right ]\right. \nonumber \\&+&\left.
\frac{1}{2} \left[\frac{r}{\sigma_X^2} - \ln \left (\frac{r}{\sigma_X^2}
\right ) - 1 \right ] - \ln(\cos\alpha)
\right\}, \label{eq:tocho_theo}
  \end{eqnarray}
where
\begin{eqnarray}
  T_1(r, \alpha, \bm v) \triangleq
(\sqrt{r} \sin\alpha + v_1)^2
\left (\frac{1}{\cos^2\beta} - 1\right)
-(\sqrt{r}\cos\alpha + v_2)^2
-v_3^2.\nonumber
\end{eqnarray}
\end{theorem}

\noindent {\it Proof.} For convenience, let us apply the
Gram--Schmidt orthogonalization procedure to the vectors $\bm u$,
$\bm x$ and $\bm w$, and then select the remaining $n-3$ orthonormal
basis functions for $\reals^n$ in an arbitrary manner. After
transforming to the resulting coordinate system, the above vectors
have the forms $\bm u = (\sqrt{n},0, 0,\ldots,0)$, $\bm x = (x_1,
x_2,0,\ldots,0)$, $\bm w = (w_1, w_2, w_3, 0, \ldots,0)$ and $\bm y
= (x_1 + w_1, x_2 + w_2, w_3, 0,\ldots,0)$, while all the components
of the noise sequence $\bm z$ will remain, in general, non--null.
From (\ref{eq:rho-det}), the false--negative event occurs whenever
\begin{eqnarray}
  \frac{(x_1 + w_1 + z_1)^2}{(x_1 + w_1 + z_1)^2 +
    (x_2 + w_2 + z_2)^2 + (w_3 + z_3)^2 + \sum_{j=4}^n z_j^2}
  < \cos^2\beta, \nonumber
\end{eqnarray}
where $w_1^2 + w_2^2 + w_3^2 \leq n D$,
$x_1^2 = nr\sin^2\alpha$ and
$x_2^2 = nr\cos^2\alpha$,
with $r$ being given by $r \triangleq \frac{||\bm x||^2}{n}$, and $\alpha \triangleq
\arcsin \left (\frac{< \bm x, \bm u>}{||\bm x|| \cdot
    ||\bm u|| } \right )$. Equivalently, the false negative event can be rewritten as:
\begin{eqnarray}
  &&(x_1 + \sqrt{n}v_1 + z_1)^2 \left (\frac{1}{\cos^2(\beta)} - 1 \right )
  - (x_2 + \sqrt{n} v_2 + z_2)^2 - (\sqrt{n} v_3 + z_3)^2 \nonumber \\
  & = & (\sqrt{nr} \sin(\alpha) + \sqrt{n}v_1 + z_1)^2
  \left (\frac{1}{\cos^2(\beta)} - 1 \right )\nonumber\\
  &-& \left [\sqrt{nr}\cos(\alpha) + \sqrt{n}v_2 +
    z_2\right ]^2
  - (\sqrt{n}v_3 + z_3)^2
   <  \sum_{j=4}^n z_j^2 =
  (n-3)q, \nonumber
\end{eqnarray}
where $q \triangleq \frac{1}{n-3} \sum_{j=4}^n z_j^2$. By defining
\begin{eqnarray}
  T_1 \triangleq
(\sqrt{r} \sin\alpha + v_1)^2
  \left (\frac{1}{\cos^2\beta} - 1 \right )
  - (\sqrt{r}\cos\alpha + v_2)^2
  - v_3^2,
\label{eq:T1}
\end{eqnarray}
and
\begin{eqnarray}
  T_2 &\triangleq& - [z_1^2 + 2 z_1 (\sqrt{nr}\sin\alpha + \sqrt{n} v_1)]
  \left (\frac{1}{\cos^2\beta} - 1 \right ) + z_2^2 \nonumber \\
  &+&2 z_2 \left [\sqrt{nr} \cos\alpha + \sqrt{n} v_2 \right ] +
  z_3^2 + 2 \sqrt{n} v_3 z_3, \nonumber
\end{eqnarray}
the presentation of the false negative event can be further modified to
\begin{eqnarray}
  n T_1 < (n-3)q + T_2, \nonumber
\end{eqnarray}
or equivalently
\begin{eqnarray}
  q > \frac{n T_1}{n-3} - \frac{T_2}{n-3}. \nonumber
\end{eqnarray}
Next, observe that $\frac{(n-3)q}{\sigma_Z^2}$ is a $\chi^2$
random variable with $n-3$ degrees of freedom, i.e.,
\begin{eqnarray}
  f_{Q}(q) = \left \{ \begin{array}{ll} \frac{n-3}{\sigma_Z^2}
      \left (\frac{1}{2}\right)^{(n-3)/2} \frac{1}{\Gamma\left (\frac{n-3}{2}
    \right )}
  \left (\frac{(n-3)q}{\sigma_Z^2} \right )^{\left ( \frac{n-3}{2} -1
      \right )} e^{-\frac{(n-3)q}{2\sigma_Z^2}}, & \quad \quad \textrm{  if  } q\geq0\\
    0, & \quad \quad \textrm{  elsewhere} \end{array}\right ..
\end{eqnarray}
By the same token, $R = \frac{|| \bm X ||^2}{n}$,
is also a $\chi^2$ distribution, this time, with $n$ degrees of freedom, and so
its density is given by
\begin{eqnarray}
f_R(r) = \left \{ \begin{array}{ll} \frac{n}{\sigma_X^2}\left (\frac{1}{2} \right )^{n/2}
    \frac{1}{\Gamma \left (\frac{n}{2}\right)} \left ( \frac{n r}{\sigma_X^2}
      \right )^{\left (\frac{n}{2} - 1 \right )} e^{-\frac{n r}{2\sigma_X^2}},
      & \quad \quad \textrm{  if  } r\geq0\\
      0, & \quad \quad \textrm{  elsewhere} \end{array}\right ..
\end{eqnarray}
Defining $\Psi=\arcsin(<\bm X,\bm u>/\|\bm X\|)$,
we have (in the absence of a watemark):
\begin{eqnarray}
  P(\Psi \leq \alpha) = 1 - \frac{A_n(\pi/2 - \alpha)}{2 A_n(\pi/2)}, \nonumber
\end{eqnarray}
from which it follows that the pdf of $\Psi$ is
\begin{eqnarray}
  f_\Psi(\alpha) = \frac{\partial P(\Psi \leq \alpha)}{\partial \alpha}
  = \frac{2\Gamma \left (\frac{n}{2} \right )}
  { \sqrt{\pi}\Gamma\left (\frac{n-1}{2}\right )}
  \cos^{n-2}\alpha. \nonumber
\end{eqnarray}
and so
\begin{eqnarray}
  P_{fn} &=& \int_{\alpha = -\pi/2}^{\pi/2} \int_{r=0}^{+\infty}
  \int_{z_3 = - \infty}^{+\infty}   \int_{z_2 = - \infty}^{+\infty}
  \int_{z_1 = - \infty}^{+\infty} \int_{q=\max(0, \frac{n T_1}{n-3} - \frac{T_2}{n-3})}^{+\infty}
  \frac{n-3}{\sigma_Z^2}\left (\frac{1}{2}\right)^{(n-3)/2} \nonumber \\ &&\frac{1}{\Gamma\left (\frac{n-3}{2}
    \right )} \left (\frac{(n-3)q}{\sigma_Z^2} \right )^{\left ( \frac{n-3}{2} -1
      \right )} e^{-\frac{(n-3)q}{2\sigma_Z^2}}
    \frac{e^{-\frac{z_1^2}{2\sigma_Z^2}}}{\sqrt{2 \pi \sigma_Z^2}}
    \frac{e^{-\frac{z_2^2}{2\sigma_Z^2}}}{\sqrt{2 \pi \sigma_Z^2}}
    \frac{e^{-\frac{z_3^2}{2\sigma_Z^2}}}{\sqrt{2 \pi \sigma_Z^2}}\nonumber \\
    && \frac{n}{\sigma_X^2}\left (\frac{1}{2} \right )^{n/2}
    \frac{1}{\Gamma \left (\frac{n}{2}\right)} \left ( \frac{n r}{\sigma_X^2}
      \right )^{\left (\frac{n}{2} - 1 \right )} e^{-\frac{n r}{2\sigma_X^2}}
      \frac{2\Gamma \left (\frac{n}{2} \right )}
      { \sqrt{\pi}\Gamma\left (\frac{n-1}{2}\right )}
      \cos^{n-2}\alpha\cdot
      dq dz_1 dz_2 dz_3 dr d\alpha\nonumber .
\end{eqnarray}
Using the facts that
$\lim_{n\rightarrow \infty}\frac{nT_1}{n-3} - \frac{T_2}{n-3} =
T_1$ and that $T_2$ grows sublinearly with $n$, we get
\begin{eqnarray}
  \lim_{n \rightarrow \infty} -\frac{1}{n} \ln P_{fn}
& = & -\frac{1}{2} -\frac{1}{2}  - \lim_{n \rightarrow \infty} \frac{1}{n} \ln \int_{\alpha =
-\pi/2}^{\pi/2} \int_{r=0}^{+\infty}
  \int_{z_3 = - \infty}^{+\infty}   \int_{z_2 = - \infty}^{+\infty}
  \int_{z_1 = - \infty}^{+\infty} \int_{q=\max(0, T_1)}^{+\infty}
  \nonumber \\
  && \frac{e^{-\frac{z_1^2}{2\sigma_Z^2}}}{\sqrt{2 \pi \sigma_Z^2}}
    \frac{e^{-\frac{z_2^2}{2\sigma_Z^2}}}{\sqrt{2 \pi \sigma_Z^2}}
    \frac{e^{-\frac{z_3^2}{2\sigma_Z^2}}}{\sqrt{2 \pi \sigma_Z^2}}\times \nonumber \\
&& e^{( \frac{n-3}{2} -1) \ln(\frac{q}{\sigma_Z^2}) }
      e^{-\frac{(n-3)q}{2\sigma_Z^2}}
      e^{\left (\frac{n}{2} - 1 \right ) \ln \frac{r}{\sigma_X^2} } e^{-\frac{n
      r}{2\sigma_X^2}}\times \nonumber \\
&& e^{(n-2)\ln(\cos\alpha)}
      dq dz_1 dz_2 dz_3 dr d\alpha\nonumber .
\end{eqnarray}
where we used the fact that
\begin{equation}
   \lim_{n \rightarrow \infty} \frac{1}{n}\ln \left[ \frac{(1/2)^{\frac{n}{2}}
   n^{\frac{n-2}{2}}}{\Gamma(n/2)}\right] = \frac{1}{2}.
\end{equation}
Finally, by using the saddle--point method
\cite{Wong}, the exponential rate of
this multi--dimensional integral is dominated by the point at which
the integrand is maximum, and
we obtain the result asserted in the theorem.
This completes the proof of Theorem 1.

\section{The Optimum Watermarking Displacement Signal}\label{sec:opt_emb}

Having derived $E_{fn}$
as a function of $\bm v$, we are now ready to derive the
main result of this paper, which is the optimum
embedding function, i.e., the one that maximizes $E_{fn}$.

\begin{theorem}\label{th:teorema2}
The maximum false--negative exponent, $E_{fn}$, subject to the constraint $v_1^2+v_2^2+v_3^2\le D$,
is achieved by $v^*=(v_1^*,v_2^*,v_3^*)$ where:
\begin{eqnarray}
  v_1^* &=& \pm \sqrt{D - r \cos^4\beta},
\nonumber \\
  v_2^* &=& - \sqrt{r}\cos^2\beta, \nonumber\\
  v_3^* &=& 0. \nonumber
\end{eqnarray}
\end{theorem}

\noindent {\it Proof.} Consider first the dependence of $E_{fn}$ on
$\alpha$. On the one hand, $-\ln(\cos\alpha)$ is minimized when
$\alpha = 0$. On the other hand, $T_1$ also depends on $\alpha$.
Since $E_{fn}$ is monotonically non--decreasing in $T_1$ and the
distortion is insensitive to the sign of any component of the
watermark, it is seen from eq.\ (\ref{eq:T1}) that the signs $v_1$
and $v_2$ should be such that $v_1\sin\alpha \geq 0$, and $v_2
\cos\alpha \leq 0$. Therefore $T_1(r, \alpha)$ is even in $\alpha$,
and its minimum is reached at $\alpha = 0$. This means that the
minimum of (\ref{eq:tocho_theo}) is obtained for $\alpha = 0$, and
then (\ref{eq:tocho_theo}) can be rewritten as
\begin{eqnarray}
  \lim_{n \rightarrow \infty} -\frac{1}{n} \ln P_{fn} &=&
  \min_{(q, r) \in [\max(0, T_1(r)), \infty)\times [0, \infty)}
  \frac{1}{2} \left[\frac{q}{\sigma_Z^2} - \ln \left (\frac{q}{\sigma_Z^2}
      \right ) - 1 \right ] \nonumber \\&+&
    \frac{1}{2} \left[\frac{r}{\sigma_X^2} - \ln \left (\frac{r}{\sigma_X^2}
      \right ) - 1 \right ]. \label{eq:tocho2}
\end{eqnarray}
As the objective function is convex in $(r,
q)$, and the global minimum is at $(\sigma_X^2, \sigma_Z^2)$, the minimum in
(\ref{eq:tocho2}) would vanish if $(\sigma_Z^2, \sigma_X^2)
\in [\max(0, T_1(r)), \infty)\times [0, \infty)$. Otherwise,
the minimum lies on the boundary, i.e., it
is a point of the form $(T_1(r), r)$, with $r
\geq 0$.

Consider next the optimization of
$(v_1, v_2, v_3)$.
Observe that the only influence of
$\bm v$ on $E_{fn}$ is via $T_1$.
Thus, $\bm v$ should be chosen so as to maximize
$T_1$.
Given that $\alpha = 0$, $T_1$
can be written as
\begin{eqnarray}
  T_1 =
  v_1^2
  \left (\frac{1}{\cos^2\beta} - 1 \right )
  - (\sqrt{r} + v_2)^2
  - v_3^2, \nonumber
\end{eqnarray}
which should be maximized over $\bm v$ subject to
\begin{eqnarray}
  v_1^2 + v_2^2 + v_3^2 \leq D. \nonumber
\end{eqnarray}
Obviously any non--zero value of $v_3$, both decreases $T_1$ and
reduces the distortion budget remaining for $v_1$ and $v_2$. Thus,
$v_3^* = 0$. Now, $T_1$ is monotonically increasing in $v_1^2$, so
the maximum must be achieved for $v_1^2 + v_2^2 = D$, which enables
to express $T_1$ as\footnote{Note that two solutions are possible
for $v_2$, namely $v_2 = \pm \sqrt{D - v_1^2}$. Here we take the
negative one, since, as we noted before, $v_2$ and $cos\alpha$ must
have opposite signs and $-\pi/2 \le \alpha \le \pi/2$, thus
$cos\alpha$ is always positive.}
\begin{eqnarray}
  T_1 =   v_1^2
  \left (\frac{1}{\cos^2\beta} - 1 \right )
  - \left [\sqrt{r} - \sqrt{D - v_1^2}\right ]^2. \nonumber
\end{eqnarray}
Equating $dT_1/dv_1$ to zero and solving for $v_1$, we obtain three solutions:
\begin{eqnarray}
  \left \{
    \begin{array}{l}
      v_1 = 0 \\
      v_1 = -\sqrt{D - r \cos^4\beta}\\
      v_1 = \sqrt{D - r \cos^4\beta}
    \end{array}
  \right. . \nonumber
\end{eqnarray}
Considering the second derivative, it is easy to see that for $v_1^*
= \pm \sqrt{D - r \cos^4\beta}$ one obtains
maxima of $T_1$, yielding $v_2^*
= -\sqrt{r}\cos^2\beta$, and a corresponding value of $T_1 = D
\tan^2\beta - r\sin^2\beta$.

\subsection{Discussion}

First, observe that the watermarking displacement signal $\bm w$, and
therefore also the watermarked sequence $\bm y$, lies in the plane spanned by the
watermark $\bm u$ and the host signal $\bm x$ (a similar conclusion
was reached in \cite{Merhav08} in the attack--free
case). This allows to express the optimum watermarking displacement signal, as
well as the watermarked sequence, as a combination of the host
signal and the watermark, leading to the following result:

\begin{corollary}\label{th:corolario1}
  The optimum watermarked signal is given by $\bm y = a \bm x + b \bm u$,
  where
  \begin{eqnarray}
    a &= & 1 - \frac{\cos^2\beta}{\cos\alpha}, \nonumber\\
    b &=& \sqrt{r} \cdot \tan\alpha \cos^2 \beta \pm \sqrt{D -
    r \cos^4\beta}. \nonumber
  \end{eqnarray}
\end{corollary}

\noindent
{\it Proof.}
From Theorem~\ref{th:teorema2}, we have:
\begin{eqnarray}
  y_1 &=& \sqrt{n r} \sin\alpha \pm \sqrt{n(D - r \cos^4\beta)}
  \label{eq:y_1_corolario} \\
  y_2 &= & \sqrt{n r} [\cos\alpha - \cos^2\beta]. \nonumber
\end{eqnarray}
On the other hand, $y_2 = a\sqrt{n r}\cos\alpha$, and so, we can
conclude that $a = 1 - \frac{\cos^2\beta}{\cos\alpha}$. To find $b$,
we use $y_1 = a \sqrt{nr}\sin\alpha + b \sqrt{n},$ which when
combined with (\ref{eq:y_1_corolario}), gives the value of $b$ is
asserted in Corollary 1. This completes the proof of Corollary 1.

It should also be pointed out that the optimum embedding strategy
depends neither on $\sigma_X^2$ nor on
$\sigma_Z^2$, which is the desirable required universality feature.
As a consequence, the embedding
strategy is the same for the attack--free case, studied in detail
in Section~\ref{sec:noiseless}.

\begin{figure}[t]
  \begin{center}
    \psfrag{a}[c][]{$\alpha$}
    \psfrag{b}[c][]{$\beta$}
    \psfrag{c}[c][]{$\bm x$}
    \psfrag{d}[c][]{$\bm w_{opt}^{min}$}
    \psfrag{e}[l][]{$\bm w_{se}^{min}$}
    \psfrag{f}[c][]{$\bm y_{opt}^{min}$}
    \psfrag{g}[l][]{$\bm y_{se}^{min}$}
    \psfrag{h}[h][]{$\bm y_{opt}^{rob}$}
    \psfrag{i}[h][]{$\bm y_{se}^{rob}$}
    \psfrag{z}[c][]{$\bm u$}
    \includegraphics[width=0.6\linewidth]{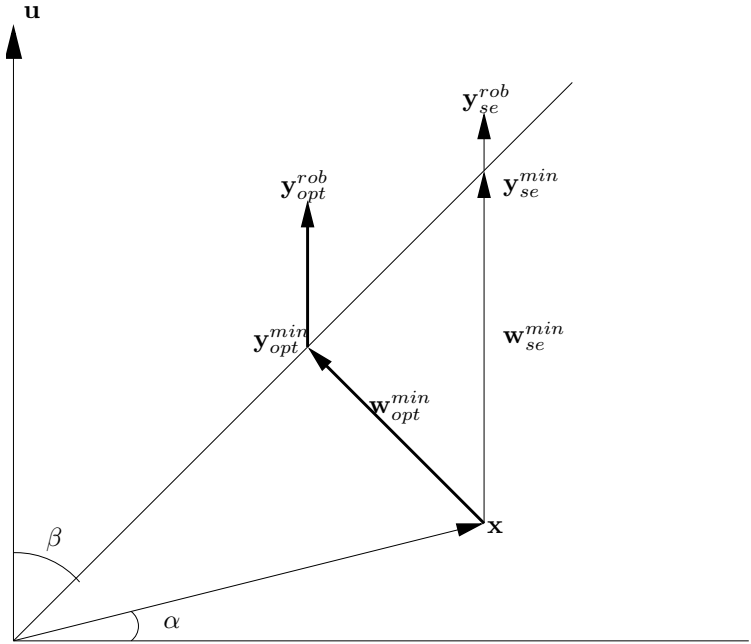}
    \caption{Geometrical interpretation of the optimum embedding problem, and comparison
    between the sign-embedder and the optimum embedder. $\bm w_{opt}^{min}$
  and $\bm w_{se}^{min}$ denote the minimum norm watermarking displacement signals that
  produce signals in the detection region, for both the optimal embedder and the
  sign embedder, respectively. The corresponding watermarked signals
  are $\bm y_{opt}^{min}$ and $\bm y_{se}^{min}$. Furthermore, one can see the
  watermarked signals for the optimal embedder and the sign embedder
  when part of the embedding distortion can be used to gain some
  robustness to noise (denoted by $\bm y_{opt}^{rob}$ and $\bm y_{se}^{rob}$).}\label{fig:triang}
  \end{center}
\end{figure}

The geometrical interpretation of the embedding strategy is
the following: the embedder devotes part of the allowed
distortion budget to scale down the host signal, thus reducing its
interference, and then injects the remaining energy in the
direction of the watermark. In fact, this explains why only the
component of the watermarked signal in the direction of the
watermark (i.e., $b$) depends on
$D$.
For illustration, we compare the optimum embedding and
the sign-embedder introduced in \cite{Merhav08}. For the sign
embedder, the watermarked signal is given by $\bm y_{se} = \bm x +
\textrm{sign}(\bm x^t\cdot \bm u)\sqrt{D}\bm u$, so the
watermarking displacement signal can be written as $\bm w_{se} =
\textrm{sign}(\bm x^t\cdot \bm u)\sqrt{D}\bm u$. The two
strategies are compared in Fig.~\ref{fig:triang}, where it is easy
to see that the proposed strategy is that of minimizing the
embedding distortion necessary for obtaining a watermarked signal.
It is also interesting to observe that the optimum embedding
technique given by Theorem 2, could not be described by
\cite{Furon07}, as in that case the watermarking displacement signal
direction is just a function of the host signal, and it is scaled
for obtaining the desired distortion.

Another way to look at Theorem 2 is by
evaluating a joint condition on the embedding distortion and the false--positive
exponent (or equivalently on $\beta$) that allows to
obtain a false--negative error exponents: if $T_1 \leq 0$,
then the optimization in (\ref{eq:tocho2}) is performed on the
region $[0, \infty)\times [0, \infty)$, so any pair $(\sigma_Z^2,
\sigma_X^2)$, even with $\sigma_Z^2 = 0$, will be in the allowed
region, yielding a vanishing error exponent. The condition that
permits to avoid this situation is $r \leq
\frac{D}{\cos^2\beta}$. We can reach the same result by
considering the case $\alpha = 0$, which is the case that
captures most of probability. In this case, the two
components of the watermarked signal $\bm y$ are given by
\begin{eqnarray}
  y_1 &=& \pm \sqrt{n(D - r \cos^4\beta)}, \nonumber \\
  y_2 &=& \sqrt{nr}(1-\cos^2\beta), \nonumber
\end{eqnarray}
or equivalently $a=1 - \cos^2\beta$ and $b = \pm \sqrt{D - r
\cos^4\beta}$. Therefore, when $D = r \cos^2\beta$ the
watermarked signal is the intersection of the boundary of the
detection region and the perpendicular vector to that boundary that
goes through $\bm x$. On the other hand, when $D < r
\cos^2\beta$, even in the noiseless case, one cannot ensure that
the embedding distortion constraint allows to produce a signal in
the detection region, so the embedding function in that case will
not be so important. In fact, regardless of the embedding
function we choose, the false negative error exponent would vanish.

\subsection{False Negative Exponent of the Optimum Embedder}

Having solved the optimum embedding problem, we can compute the
false--negative exponent achieved by the optimum embedder and
compare it to previous results in the literature. To
do so, the optimization in (\ref{eq:tocho2}) is performed over
points of the form $(T_1(r), r)= (D \tan^2\beta -
r\sin^2\beta, r)$, with $0 \leq r \leq \frac{D}{\cos^2\beta}$.
The derivative of (\ref{eq:tocho2}) with respect to $r$ takes the
value
\begin{eqnarray}
  \frac{1}{2}\left ( - \frac{1}{r} + \frac{1}{\sigma_X^2}
    + \frac{\cos^2\beta}{D - r \cos^2\beta} - \frac{\sin^2\beta}{\sigma_Z^2}\right ), \nonumber
\end{eqnarray}
which is piecewise convex in $(0,
D/\cos^2\beta)$, and $(D/\cos^2\beta, \infty)$. Due to the
constraints introduced previously, we are interested in the minimum
in the interval $(0, D/\cos^2\beta)$, which is achieved when
\begin{eqnarray}
  r^* &=& \bigg ( D \sigma_Z^2 + 2\sigma_Z^2\sigma_X^2 \cos^2\beta-
    D \sigma_X^2 \sin^2\beta  \nonumber \\ &&-  \sqrt{D^2 \sigma_Z^4 + 4
      \sigma_Z^4 \sigma_X^4 \cos^4\beta - 2D^2\sigma_Z^2\sigma_X^2
      \sin^2\beta+ D^2\sigma_X^4\sin^4\beta} \bigg )\times \nonumber \\
    && \bigg [2(\sigma_Z^2
    \cos^2\beta - \sigma_X^2\cos^2\beta\sin^2\beta)\bigg ]^{-1}. \label{eq:r*}
\end{eqnarray}
By replacing $r$ with $r^*$ in the definition of $T_1(r)$ we get the
value of $q^*$, then we insert $r^*$ and $q^*$ in (\ref{eq:tocho2}),
and finally obtain the optimum error exponent for the AWGN case:
\begin{eqnarray}
  q^* &=& \Bigg [\bigg ( 2D \sigma_Z^2 + \sqrt{16 \sigma_Z^4 \sigma_X^4
    \cos^4\beta + D^2\left [2 \sigma_Z^2 - \sigma_X^2(1 - \cos(2\beta))
    \right]^2 } \bigg) \tan^2\beta \nonumber \\
  &-&2 \sigma_X^2 \sin^2\beta \left ( 2 \sigma_Z^2 + D\tan^2\beta
  \right ) \Bigg] \left [ 4 \left (\sigma_Z^2 - \sigma_X^2 \sin^2\beta
    \right ) \right]^{-1}, \label{eq:q*} \\
  E_{fn}^*&=&\frac{1}{2} \left[\frac{q^*}{\sigma_Z^2} - \ln \left (\frac{q^*}{\sigma_Z^2}
    \right ) - 1 \right ] +
  \frac{1}{2} \left[\frac{r^*}{\sigma_X^2} - \ln \left (\frac{r^*}{\sigma_X^2}
    \right ) - 1 \right ]. \label{eq:E_fn*}
\end{eqnarray}
Note that due to the choice of ${\cal U}$ and the symmetry of the
Gaussian distribution followed by the host around zero, the
false-negative error exponent does not depend on the particular
choice of the watermark $\bm u$.

\begin{figure}[t]
  \begin{center}
    \includegraphics[width=0.85\linewidth]{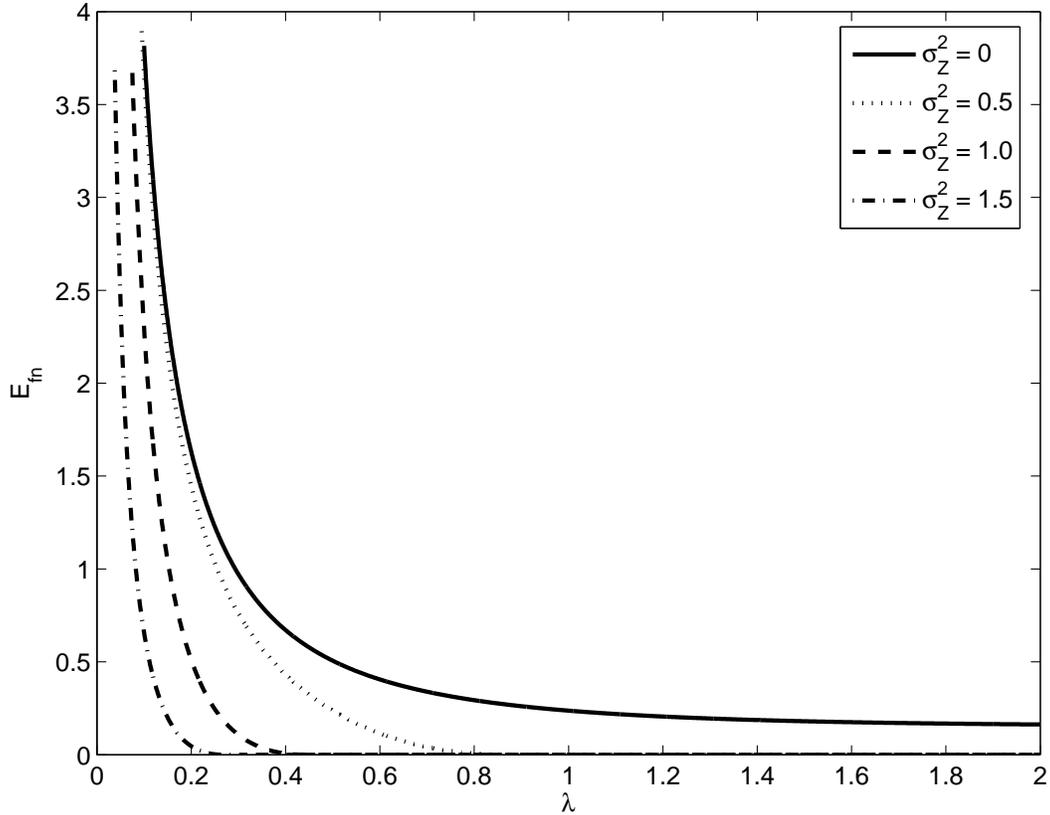}
    \caption{False negative error exponent as a function
      of $\lambda$, for several powers of AWGN. $\sigma_X^2 = 1$ and $D=2$.}
    \label{fig:Efn}
  \end{center}
\end{figure}

\begin{figure}[t]
  \begin{center}
    \includegraphics[width=0.85\linewidth]{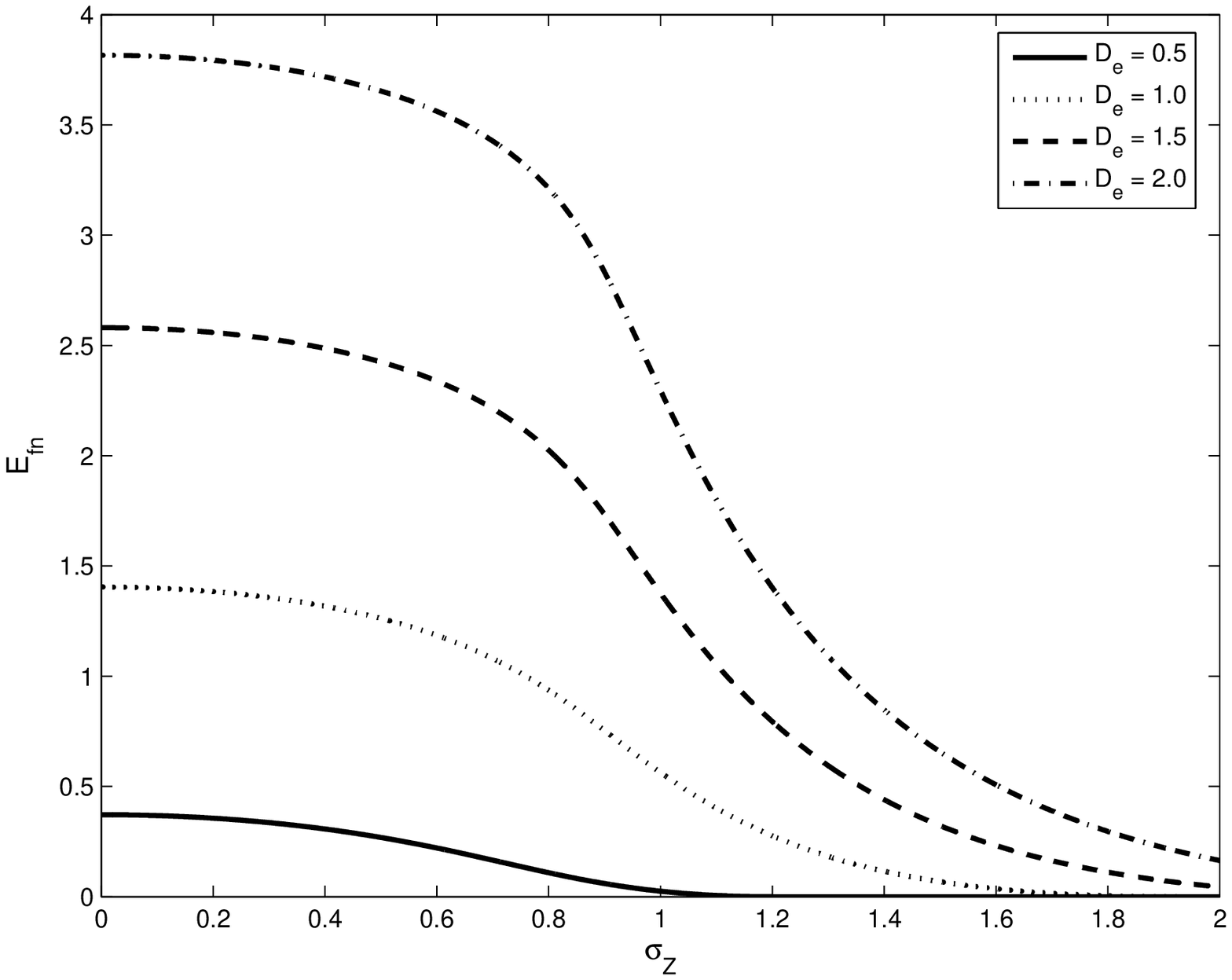}
    \caption{False negative error exponent  as a function
      of $\sigma_Z$, for several embedding distortions.
      $\sigma_X^2 = 1$ and $\lambda=0.1$.}
    \label{fig:Efn_vs_sigmaN}
  \end{center}
\end{figure}

\begin{figure}[t]
  \begin{center}
    \includegraphics[width=0.85\linewidth]{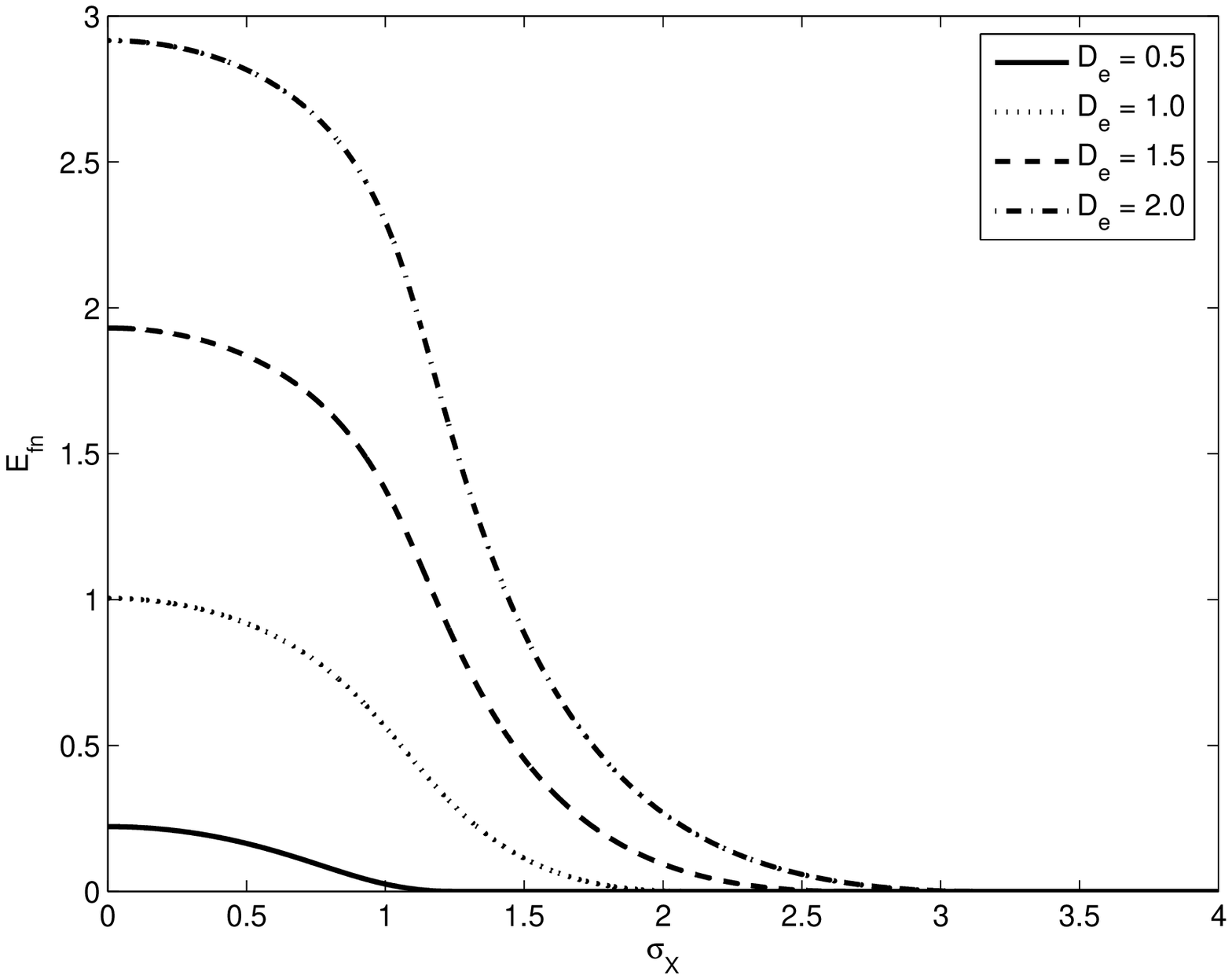}
    \caption{False negative error exponent as a function
      of $\sigma_X$, for several embedding distortions.
      $\sigma_Z^2 = 1$ and $\lambda=0.1$.}
    \label{fig:Efn_vs_sigmaX}
  \end{center}
\end{figure}

In Figs.\ \ref{fig:Efn}, \ref{fig:Efn_vs_sigmaN} and
\ref{fig:Efn_vs_sigmaX} the behavior of $E_{fn}^*$ is depicted as a function of
various parameters. As expected, the
false--negative exponent decreases when the false--positive
exponent $\lambda$, the attack variance
$\sigma_Z^2$, or the host variance $\sigma_X^2$, increase,
while it increases with $D$.

\subsection{Numerical Results}

\begin{figure}[t]
  \begin{center}
    \includegraphics[width=0.85\linewidth]{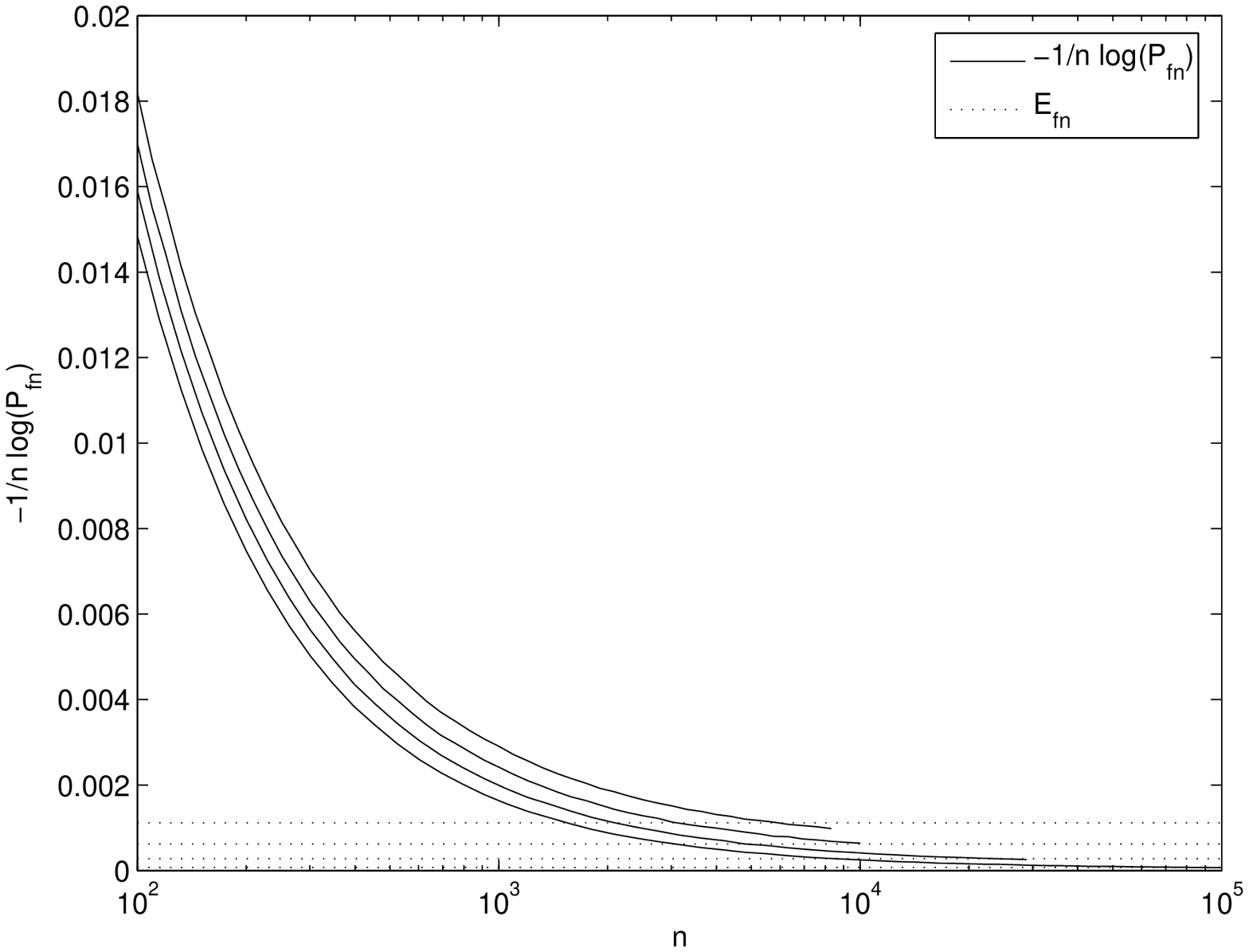}
    \caption{Theoretical false negative error exponent and  $-\frac{1}{n} \log(P_{fn})$ as a function
      of the number of dimensions $n$. $D=2$, $\sigma_X^2 = 1$
      and $\lambda=0.6$, and $\sigma_Z^2$ equal to $0.52$, $0.53$, $0.54$ and
    $0.55$, respectively (from top to bottom).}
    \label{fig:Efn_theo_vs_emp1}
  \end{center}
\end{figure}

\begin{figure}[t]
  \begin{center}
    \includegraphics[width=0.85\linewidth]{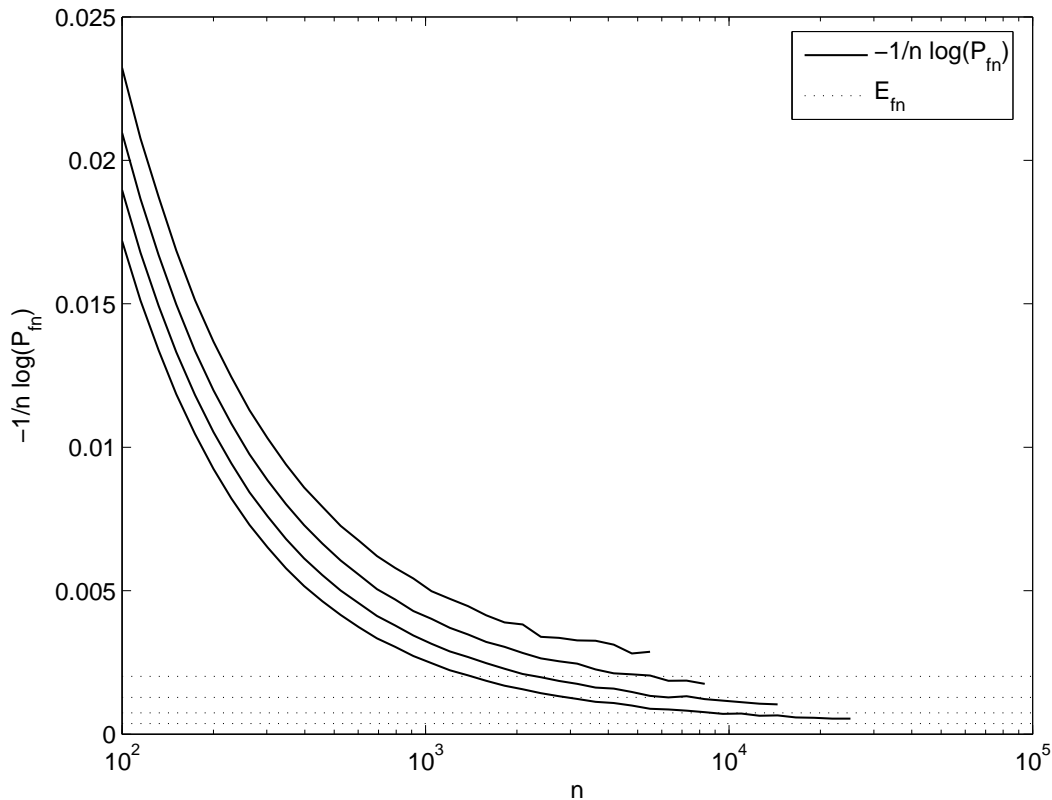}
    \caption{Theoretical false negative error exponent and  $-\frac{1}{n} \log(P_{fn})$ as a function
      of the number of dimensions $n$. $D=0.75$, $\sigma_X^2 = 1$
      and $\sigma_Z^2=0$, and $\lambda$ equal to $0.58$, $0.6$, $0.62$ and
    $0.64$, respectively (from top to bottom).}
    \label{fig:Efn_theo_vs_emp2}
  \end{center}
\end{figure}

In order to validate the theoretical results with numerical ones, we
compare the false--negative exponent with the empirical values of
$-\frac{1}{n} \ln P_{fn}$, for large $n$. Although large values of
$n$ and $  -\frac{1}{n} \log(P_{fn})$ can not be considered
simultaneously, due to the resulting very small probability of false
negative, in Fig.~\ref{fig:Efn_theo_vs_emp1} we can see the
similarity between $E_{fn}^*$ and its empirical approximation when
$n$ increases, for different values of $\sigma_Z^2$. Furthermore, in
Fig.~\ref{fig:Efn_theo_vs_emp2} we compare the empirical
approximation of the false--negative exponent to its theoretical
value for the attack--free case (special attention will be paid to
this particular case in Section~\ref{sec:noiseless}), for different
values of $\lambda$. As expected, the larger is $\lambda$, the
smaller is the false--negative exponent.

\section{The Attack--Free Case}\label{sec:noiseless}
As a special case of Theorem 1 and
Theorem 2, we calculate the false--negative exponent for
the noiseless case ($\sigma^2_Z = 0$). By computing the limit of
$\sigma_Z^2\to 0$ in (\ref{eq:r*}), it is easy to see that in
the attack--free case, we have:
\begin{eqnarray}
  \lim_{\sigma_Z^2 \rightarrow 0} r^* &=& \frac{-2D \sigma_X^2 \sin^2\beta}
  {-2\sigma_X^2 \cos^2\beta\sin^2\beta}= \frac{D}{\cos^2\beta} =
\frac{D}{1 - e^{-2\lambda}}.\label{eq:r*_noiseless}
\end{eqnarray}
To compute $\lim_{\sigma_Z^2 \rightarrow 0} \frac{q^*}{\sigma_Z^2}$
from (\ref{eq:q*}) we can use L'H\^opital's rule. Given that
\begin{eqnarray}
  \lim_{\sigma_Z^2 \rightarrow 0} &&\frac{\partial }{\partial \sigma_Z^2}  \Bigg [\bigg ( 2D \sigma_Z^2 + \sqrt{16 \sigma_Z^4 \sigma_X^4
    \cos^4\beta + D^2\left [2 \sigma_Z^2 - \sigma_X^2(1 - \cos(2\beta))
    \right]^2 } \bigg) \tan^2\beta \nonumber \\
  &&-2 \sigma_X^2 \sin^2\beta \left ( 2 \sigma_Z^2 + D\tan^2\beta
  \right ) \Bigg]=
  -4 \sigma_X^2\sin^2\beta, \nonumber
\end{eqnarray}
and
\begin{eqnarray}
  \lim_{\sigma_Z^2 \rightarrow 0} &&\frac{\partial }{\partial \sigma_Z^2} \sigma_Z^2
  \left [ 4 \left (\sigma_Z^2 - \sigma_X^2 \sin^2\beta
    \right ) \right] =  -4 \sigma_X^2\sin^2\beta, \nonumber
\end{eqnarray}
we conclude that
\begin{eqnarray}
  \lim_{\sigma_Z^2 \rightarrow 0} \frac{q^*}{\sigma_Z^2} &=& 1. \label{eq:q*_noiseless}
\end{eqnarray}
From (\ref{eq:r*_noiseless}) and (\ref{eq:q*_noiseless}), it is
straightforward to see that the value of the false--negative
exponent for the attack--free case is given by
\begin{eqnarray}
  \lim_{\sigma_Z^2 \rightarrow 0} E_{fn}^* &=& \left \{ \begin{array}{ll}
      0,  & \textrm{  if  } \frac{D}{1 - e^{-2\lambda}} \leq \sigma_X^2 \\
      \frac{1}{2} \left [\frac{D}{\sigma_X^2 \left (
            1 - e^{-2\lambda} \right)} - \ln \left ( \frac{D}{\sigma_X^2 \left (
              1 - e^{-2\lambda} \right)}\right) - 1\right] &
      \textrm{    elsewhere} \end{array}\right .
      . \label{eq:E_fn_noiseless}
\end{eqnarray}
In view of (\ref{eq:E_fn_noiseless}), it is interesting to note that
as long as $D> \sigma_X^2$, $E_{fn}^*>0$ for any
$\lambda$. In fact, under these conditions, the asymptotic
value of $E_{fn}$ when $\lambda\to\infty$ is
\begin{eqnarray}
  \frac{1}{2} \left [ \frac{D}{\sigma_X^2}
        - \ln{ \left ( \frac{D}{\sigma_X^2} \right ) - 1}
      \right ],
\end{eqnarray}
coinciding with the result of [2. Corollary 1].

\begin{figure}[t]
  \begin{center}
    \includegraphics[width=0.75\linewidth]{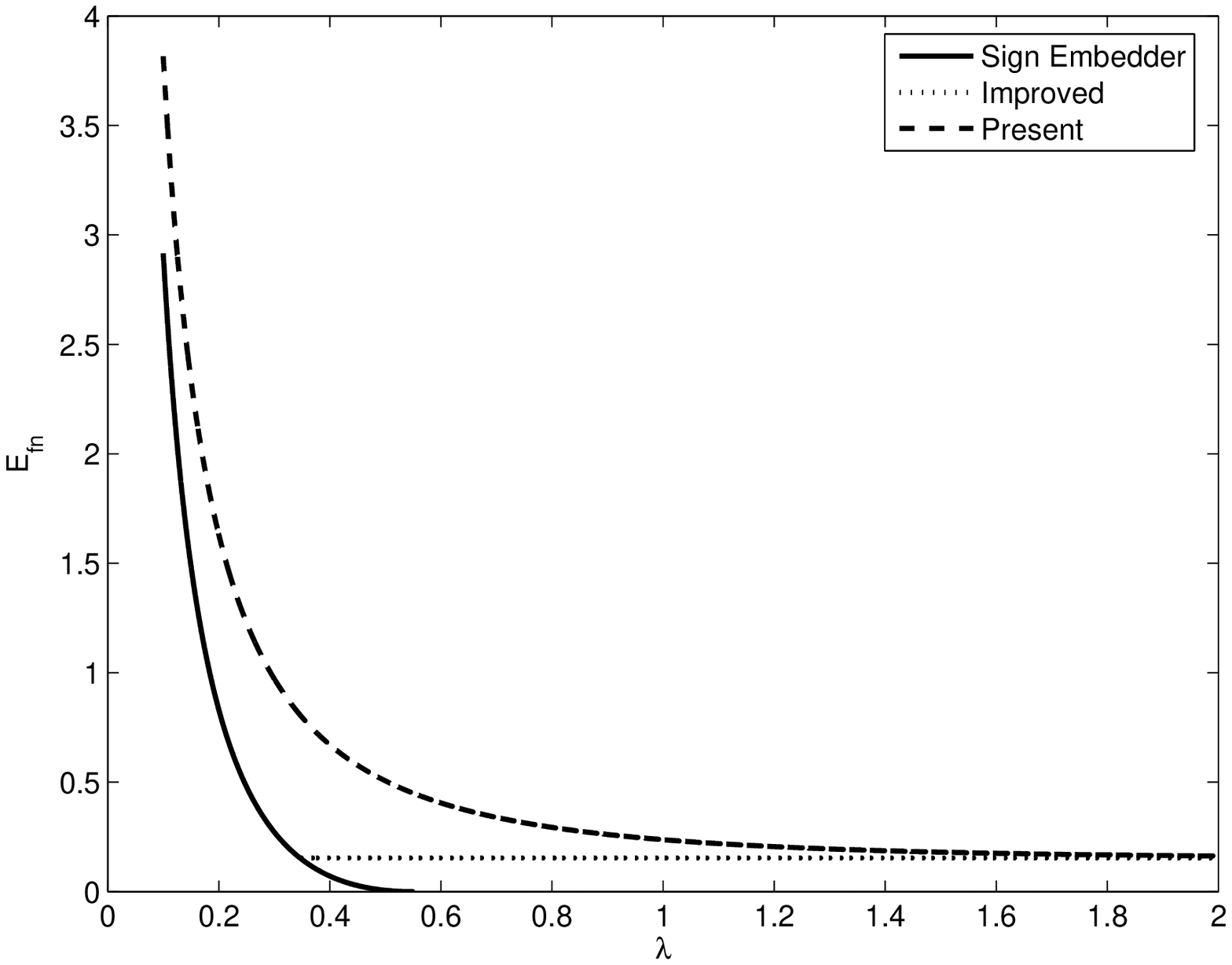}
    \caption{Comparison of the errors exponents obtained by the
      sign embedder described by Merhav and Sabbag \cite{Merhav08}, its improved version, and
      the technique presented in this work. $\sigma_X^2 = 1$ and $D=2$.}\label{old_new}
  \end{center}
\end{figure}

On the other hand, when $D \leq \sigma_X^2$ another interesting
point which reflects the goodness of the proposed strategy is the
computation of the range of values of $\lambda$ where $E_{fn}>0$ can
be achieved. In this case, the condition to be verified is
\begin{eqnarray}
  \frac{D}{1 - e^{-2\lambda}} > \sigma_X^2,
\end{eqnarray}
implying that
\begin{eqnarray}
  \lambda < - \frac{1}{2}\ln \left ( 1 - \frac{D}{\sigma_X^2}\right ) = \lambda_1,
  \textrm{   for   } D \leq \sigma_X^2,
\end{eqnarray}
whereas for the sign embedder
\cite{Merhav08}, the values of $\lambda$ for which $E_{fn} > 0$ are
those such that
\begin{eqnarray}
  \frac{D}{\sigma_X^2} > \frac{1 - e^{-2\lambda}}{e^{-2\lambda}},
\end{eqnarray}
or, equivalently,
\begin{eqnarray}
  \lambda < -\frac{1}{2}\ln \left ( \frac{\sigma_X^2}{D + \sigma_X^2} \right ) = \lambda_2,
  \textrm{   for all   } D.
\end{eqnarray}
Given that $\lambda_1 > \lambda_2$, larger values of false positive
error exponents are allowed (while still keeping $E_{fn}
> 0$) by the new embedding rule. In Figure~\ref{old_new} we compare
the bounds on the false--negative exponent for the attack--free
case found in \cite{Merhav08}, with its optimal
value derived here. As can be seen, the
improvement owing to the optimum embedding strategy is
significant, especially for small $\lambda$.

\section{Conclusions}\label{sec:conclusions}

We derived a Neyman--Pearson asymptotically optimum one--bit watermarking scheme in
the Gaussian setting, when
the detector is limited to base its decisions on second order empirical statistics only.
The scenario we considered is universal in the sense
that the variance of both the host signal and the attack are not
known to the embedder and to the detector. Our main results
are simple closed--form formulas for both the
optimum embedding function and the corresponding error exponents.
The noiseless scenario can be seen as a special case,
where we can compare the false--negative
exponent achieved by the optimum scheme with the bounds derived in
\cite{Merhav08}.
Interestingly, the optimum embedder is very simple thus opening the
door to practical implementations.

This work can be extended in many interesting directions,
including
non-Gaussian settings,
more complicated attacks, like
de-synchronization attacks \cite{barnidesync,dange07},
more detailed empirical statistics gathred by the detector,
and the
introduction of security considerations in the picture
\cite{barnisec}.

\bibliographystyle{IEEEtran}
\bibliography{biblio}

\end{document}